\documentclass[aps,twocolumn,showpacs,preprintnumbers,prl,superscriptaddress]{revtex4}
\usepackage{graphicx}
\usepackage{amssymb}
\newcommand{\gtapprox}{\raisebox{-0.5ex}{$\,\stackrel{>}{\scriptstyle\sim}\,$}}

\newcommand{\R}{\mathbf{r}}

\newcommand{\UP}{n_{\uparrow}}
\newcommand{\DN}{n_{\downarrow}}

\newcommand{\be}{\begin{equation}}
\newcommand{\ee}{\end{equation}}
\newcommand{\bea}{\begin{eqnarray}}
\newcommand{\eea}{\end{eqnarray}}
\newcommand{\bean}{\begin{eqnarray*}}
\newcommand{\eean}{\end{eqnarray*}}

\begin{document}

\title{Spin-dependent gradient correction for more accurate atomization energies of molecules} 
\author{Lucian A. Constantin}
\affiliation{Center for Biomolecular Nanotechnologies @UNILE, Istituto Italiano di Tecnologia (IIT), 
Via Barsanti, 73010 Arnesano (LE), Italy}
\author{Eduardo Fabiano}
\affiliation{National Nanotechnology Laboratory (NNL), Istituto Nanoscienze-CNR,
Via per Arnesano 16, I-73100 Lecce, Italy }
\author{Fabio Della Sala}
\affiliation{National Nanotechnology Laboratory (NNL), Istituto Nanoscienze-CNR,
Via per Arnesano 16, I-73100 Lecce, Italy }
\affiliation{Center for Biomolecular Nanotechnologies @UNILE, Istituto Italiano di Tecnologia (IIT), 
Via Barsanti, 73010 Arnesano (LE), Italy}

\date{\today}

\begin{abstract}
We discuss, simplify, and improve the spin-dependent correction 
of L.A. Constantin {\it et al}, Phys. Rev. B $\mathbf{84}$, 233103, 
for atomization energies, and 
develop a density parameter
of the form $v\propto |\nabla n|/n^{10/9}$, found from
the statistical ensemble of one-electron densities. The here constructed 
exchange-correlation generalized gradient approximations (GGAs), named
zvPBEsol and zvPBEint, show a broad applicability, and a good 
accuracy for many applications, because these corrected functionals 
significantly improve the atomization and binding energies of 
molecular systems, without worsening the behavior of the 
original functionals (PBEsol and PBEint) for other properties. 
This spin-dependent correction is also applied to meta-GGA 
dynamical correlation functionals combined with 
exact-exchange; in this case a significant (about 30\%) improvement in 
atomization energies of small molecules is found. 
\end{abstract}

\maketitle

\section{Introduction}
\label{sec1}

Kohn-Sham (KS) density functional theory (DFT) is at present one of the most 
powerful computational tools in quantum chemistry, solid-state
physics, and materials science. Therefore, it is continuously subject to
intense research. In particular, the development of new improved 
exchange-correlation (XC) functionals \cite{bookdft}, which are a main
ingredient of the KS self-consistent one-particle equations \cite{KS}, 
and that account for all the quantum many-body effects beyond the 
Hatree approximation, has been for many years, and still is, the 
main topic in this field.

The construction of XC functionals can be pursued in several different ways
\cite{scusrev}, but in any case a prominent role is played by
the availability of reference systems used to target the functional
development. In this context particularly useful are model systems, which
allow to control in a simple way specific features of the physics of
electronic systems, and thus to reduce the empiricism.
Among the others we can mention the uniform electron gas \cite{bookdft}, which is the
basic model for solid-state physics and the base of all non-empirical 
semilocal XC functionals \cite{PBE,q2D} ; the Airy gas \cite{Airy} and jellium surfaces 
\cite{LK}, 
that describe the physics of simple metal surfaces and were used to develop
generalized gradient approximations (GGA) for solids 
(AM05 \cite{AM05}, PBEsol \cite{PBEsol}) and interfaces 
(PBEint \cite{PBEint,JS}); the hydrogen atom, which is a 
basic model for atomic physics and was successfully employed, through
the analysis of the XC hole, to construct accurate meta-GGA approximations
\cite{revTPSS,br,becke87}; the semiclassical many-electron neutral atom \cite{ESa84,ELCB}
that was used to obtain a modified second-order gradient expansion
for both the kinetic \cite{LCPB09} and exchange energy \cite{EB09}, 
which are at the base of accurate XC GGA for molecular 
systems (APBE \cite{APBE,mk}), 
as well kinetic energy GGAs
constructed for the embedding theory of 
weakly-interacting molecular systems \cite{apbek} .

Recently, we also proposed \cite{zeta} a constraint 
for atomization energies at the GGA level, based on a model system
constituted by a statistical ensemble of one-electron densities. 
In particular, we considered the
reference one-electron hydrogen (H), Gaussian (G), and cuspless (C)
densities  (atomic units are used throughout, i.e., $e^2=\hbar=m_e=1$)
\begin{equation}\label{e1}
n^H(r)=\frac{e^{-2r}}{\pi},\;\;
n^G(r)=\frac{e^{-r^2}}{\pi^{3/2}},\;\;n^C(r)=\frac{(1+r)e^{-r}}{32\pi},
\end{equation}
and we showed that the accuracy of a GGA XC functional to describe cohesive 
properties is directly related to the information-entropy-like 
function \cite{Jaynes}
\begin{equation}\label{e2}
I[GGA]=-\sum_{i} p_iR_i\ln(p_iR_i),\;\;\;i=H,\; G,\; \; \rm{and} \; C,
\end{equation}
where $R=|\Delta E_{xc}/E_{x}^{exact}|$ is the relative
absolute error of the XC energy of G, H, or C (for a given GGA), and $p_i=1/3$ 
are probabilities (weights) of G, H, and C appearance in a
molecular bond (using the assumption of equ-probability).
Note that this constraint for atomization energies,
was derived from an empirical observation relating
errors in the model one-electron densities to errors in the atomization
energies, of popular GGAs (see Fig. 2 of Ref. \cite{zeta}); and from the physical explanation
 that one-electron densities are simple models for simple
bonding regions, where iso-orbital regime can be significant.
(see Fig. 1 of Ref. \cite{zeta}, and the corresponding discussion.)

This constraint was used to improve the description of cohesive energies
of finite-size and periodic systems by GGA functionals of PBE-like form
\cite{PBE} through the use of a spin-dependent  XC correction 
based on the ansatz
\cite{zeta}
\begin{equation}\label{e3}
\epsilon_{xc}^{GGA}(\alpha)=\epsilon_x^{LDA}F^{GGA}(s)+\epsilon_c^{LSDA}+f_\alpha(\phi,t)H^{GGA},
\end{equation}
where $\epsilon_x^{LDA}$ and $\epsilon_c^{LDA}$ are the LDA
exchange and correlation energies per particle; $F^{GGA}(s)$ is 
the exchange enhancement factor; 
$H^{GGA}$ is the PBE-like gradient correction for the correlation;
$\phi=((1+\zeta)^{2/3}+(1-\zeta)^{2/3})/2$ is a spin-scaling factor that lays 
between 1 (for spin-unpolarized systems) and $2^{-1/3}=0.7937$ 
(for fully spin polarized systems) and 
$\zeta=(\UP-\DN)/n$ is the relative spin polarization;
$s$ and $t$ are the reduced gradients for exchange and 
correlation respectively \cite{LM,PBW}
\begin{equation}\label{e4}
s=|\nabla n|/2 k_F n,\;\;\; t=|\nabla n|/2 k_s\phi  n\ ,
\end{equation}
with $k_F=(3\pi^2n)^{1/3}$ being the local
Fermi wavevector, and $k_s=(4k_F/\pi)^{1/2}$ being the
Thomas-Fermi screening wavevector;
finally the following spin-dependent correction has been selected
\cite{zeta}: 
\begin{equation}\label{e5}
f_\alpha(\phi,t)=\phi^{\alpha t^3}\ ,
\end{equation}
 which satisfies important physical constraints \cite{zeta}.
The parameter $\alpha$ was obtained by the minimization of $I[GGA]$,
yielding the zPBEsol and zPBEint GGA functionals \cite{zeta},
having GGA=PBEsol, $\alpha=4.8$ and GGA=PBEint, $\alpha=2.4$,
respectively.

We underline that, as for most DFT functionals \cite{scusrev},
the ansatzes of Eqs. (3) and (5) were not derived directly from
exact equations/conditions, but are
constructed with the aim to fulfill all the exact
constraints for the original GGA and additionally, the
atomization energy constraint, derived in Ref. \cite{zeta}.
Thus, the spin-dependent
correction can be viewed as an ad hoc correction for
the whole XC functional, that can be applied to GGAs constructed
for solids (see Fig. 3 of Ref. [22]).

By construction the functionals keep unchanged the behavior of the 
original functionals (PBEsol and PBEint) for spin-unpolarized systems, 
and they significantly improve 
the atomization energies of molecules and solids \cite{zeta}.
In fact, the ansatz of Eqs. (\ref{e3}) and (\ref{e5})
preserves
many important constraints of the original GGAs, including the 
spin-scaling behavior of the XC functional in the slowly-varying 
density limit \cite{PO1,CA,PW1,Rasolt1,WP1,HL1}, 
and in addition enforce the atomization energy constraint.
Note also that the spin-dependent correction of Eq. (\ref{e5}) only
acts in the rapidly-varying density regime, where the exact 
spin-dependence of the correlation is not known and the PBE-like
functionals assume the spin dependence of the second-order gradient
expansion \cite{WP1,PBE}. 

Despite their good behavior, however the zPBEint and zPBEsol functionals
display some non-negligible formal drawbacks related to the form 
used for the spin-dependent correction (Eq. (\ref{e5})). In particular, because
the correction is aimed at improving the bonding description,
it should act only in the valence regions of the electron density.
On the other hand, as it will be discussed in this paper, it turns out that in atomic 
inter-shell regions $t$ can be 
large so that, even if $\phi$ is close to 1 (as in the whole core region), 
the spin-dependent correction may be (inappropriately) important. 

In this paper, we consider in more detail the construction 
proposed in Ref. \onlinecite{zeta} and develop a new 
spin-dependent ansatz, beyond the form of Eq. (\ref{e5}), which
is able to implement the constraint of our statistical set
of one-particle densities removing the previous limitation.
To this end we introduce a new density parameter more convenient
for atomization energies, and additional constraints from partially spin-polarized
one electron densities.
Moreover, using the new ansatz we demonstrate the utility of 
a spin-dependent correction beyond the GGA level, applying our
spin-dependent factor to improve the compatibility of
meta-GGA correlation functionals with the exact-exchange (EXX).

This article is organized as follow: in Section \ref{sec3} we
propose an improved ansatz for the spin-dependent correction, which
is based on a new reduced density parameter; in Section \ref{sec4}
we show the results of several GGA functionals bearing the spin-dependent
correction, for 
atomization and binding energies, as well for other properties; 
in Section \ref{sec5} we develop spin-dependent corrections for meta-GGA 
correlation functionals, that allows a better compatibility with 
the exact exchange, improving the atomization energies by about 30\%,
and finally, in Section \ref{sec6}, we summarize our conclusions.

\section{Improved ansatz for the spin-dependent correction}
\label{sec3}
Let us consider, instead of $f_\alpha(\phi,t)$ a more general form for the spin-dependent
correction factor to be used in Eq. (\ref{e3}):
\begin{equation}\label{e14}
f_{\alpha,\omega}(\zeta,v)=e^{-\alpha v^3  |\zeta|^\omega},
\end{equation}
where $\alpha\geq 0$ and $\omega > 0$ are constants which will be fixed from 
exact constraints, and $\zeta$ is the relative spin polarization ($\zeta=1$ for 
fully-spin-polarized systems, and $\zeta=0$ for spin-unpolarized systems). 
Here, $v$ is a generalized spin-independent density parameter for which we 
use the general ansatz
\begin{equation}\label{e15}
v=t\phi \left(\frac{r_s}{3}\right)^{-x} \, ,
\end{equation}
with $x$ a parameter to be fixed from one-electron density analysis.
Note that the form of Eq. (\ref{e15}) assures that $v$ can
describe the density regime with good flexibility. In particular,
when $x=0$ we have $v= t\phi$, while for $x=-1/2$ we find $v\propto s$.

The ansatz in Eq. (\ref{e14}) satisfies the following exact properties (as the ansatz in Eq. 
(\ref{e5}):

(i) At a constant density $v=0$, thus $f_{\alpha,\omega}(\zeta,0)=1$, and Eq. (\ref{e3})
correctly recovers the LDA behavior.

(ii) For a slowly-varying density, $v$ is small and thus 
$f_{\alpha,\omega}(\zeta,v)\rightarrow 1+O(|\nabla n|^3)$, i.e. with a third power in the 
gradient of the density.
In this way the  corrected GGA of Eq. (\ref{e3}), performs exactly as the 
original PBE-like correlation one ($H_{PBE}\rightarrow \beta \phi^3 t^2 \propto  |\nabla n|^2$).
 
(iii) For any spin-unpolarized system ($\zeta=0$ everywhere) or 
regions of space where $\zeta=0$, there is no 
spin-dependent correction ($f_{\alpha,\omega}(0,v)=1$), and thus  
the  corrected GGA of Eq. (\ref{e3}) recovers the original PBE-like correlation.

Thus XC-energy density in Eq.  (\ref{e3}) with the spin-dependent correction in Eq. (\ref{e14}) 
satisfies 
the {\it same} physical constrains as the original PBEsol (PBEint).
In addition the spin-dependent correction in Eq. (\ref{e14}) is constructed so 
that $f_{\alpha,\omega}(\zeta,v)< 1$
only  in rapidly-varying spin-polarized  density regime ($\zeta \ne 0$ and $v > 1$), where no
 exact constraints are know.
In this regime the total correlation energy density of our corrected functional (i.e. 
$\epsilon_c^{LSDA}+f H^{GGA}$)
will be larger (more negative) than the original PBE-like correlation and can thus 
partially compensate the 
underestimated contribution of PBEsol (PBEint) exchange functional \cite{zeta}.

Concerning the relation between the new ansatz in Eq. (\ref{e14}) and the one in Eq. (\ref{e5}), we note
that the latter 
can be rewritten as 
$\phi^{\alpha t^3}=e^{\ln(\phi)\alpha t^3}$. For 
small spin-polarizations ($\zeta\rightarrow 0$, $\phi\rightarrow 1$), 
we have $\ln(\phi)\rightarrow-|\zeta|^2/9$, and 
$v\rightarrow t\phi\rightarrow t$ for $x=0$ and $\phi\rightarrow 1$. 
Thus, the new ansatz in Eq. (\ref{e14}) is equivalent to the one 
of Eq. (\ref{e5}) with $\omega=2$ and $x=0$ ($\alpha$ is a constant).
This equivalence is trivially valid also for fully polarized systems
(as the model one-electron densities in the statistical ensemble):
in fact in this case $\phi=2^{-1/3}$ and $\zeta=\pm 1$ so that $\ln(\phi)$ is just a constant, 
$|\zeta|^\omega=1$ 
and again $v$ is proportional to $t$ for $x=0$.

The main advantage of  Eq. (\ref{e14}) with respect to Eq. (\ref{e5}) is that
the dependence of the correction factor $f$ from the spin- and 
spatial-properties of the density is decoupled and controlled respectively
by the $\omega$ factor and the density parameter $v$. Therefore,
a fine tuning of the correction is possible to make it relevant only in 
valence and bonding regions.

To fix the parameters in Eqs. (\ref{e14}) and (\ref{e15}), we note
that for fully-spin-polarized systems 
$f(v,\zeta=1)=e^{-\alpha v^3}$, thus the parameters $\alpha$ and $x$ 
can be found from the one-electron-densities statistical ensemble, 
by minimization of the information-entropy-like function. 
To perform such minimization, the form of the GGA functional 
entering Eq. (\ref{e3}) must be fixed. As in Ref. \onlinecite{zeta},
we consider in the present case GGA=PBEint and GGA=PBEsol, obtaining
$x=1/6$ and $\alpha=1.0$ and $\alpha=1.8$ for PBEint and PBEsol,
respectively.
 
%
\begin{figure}
\includegraphics[width=\columnwidth]{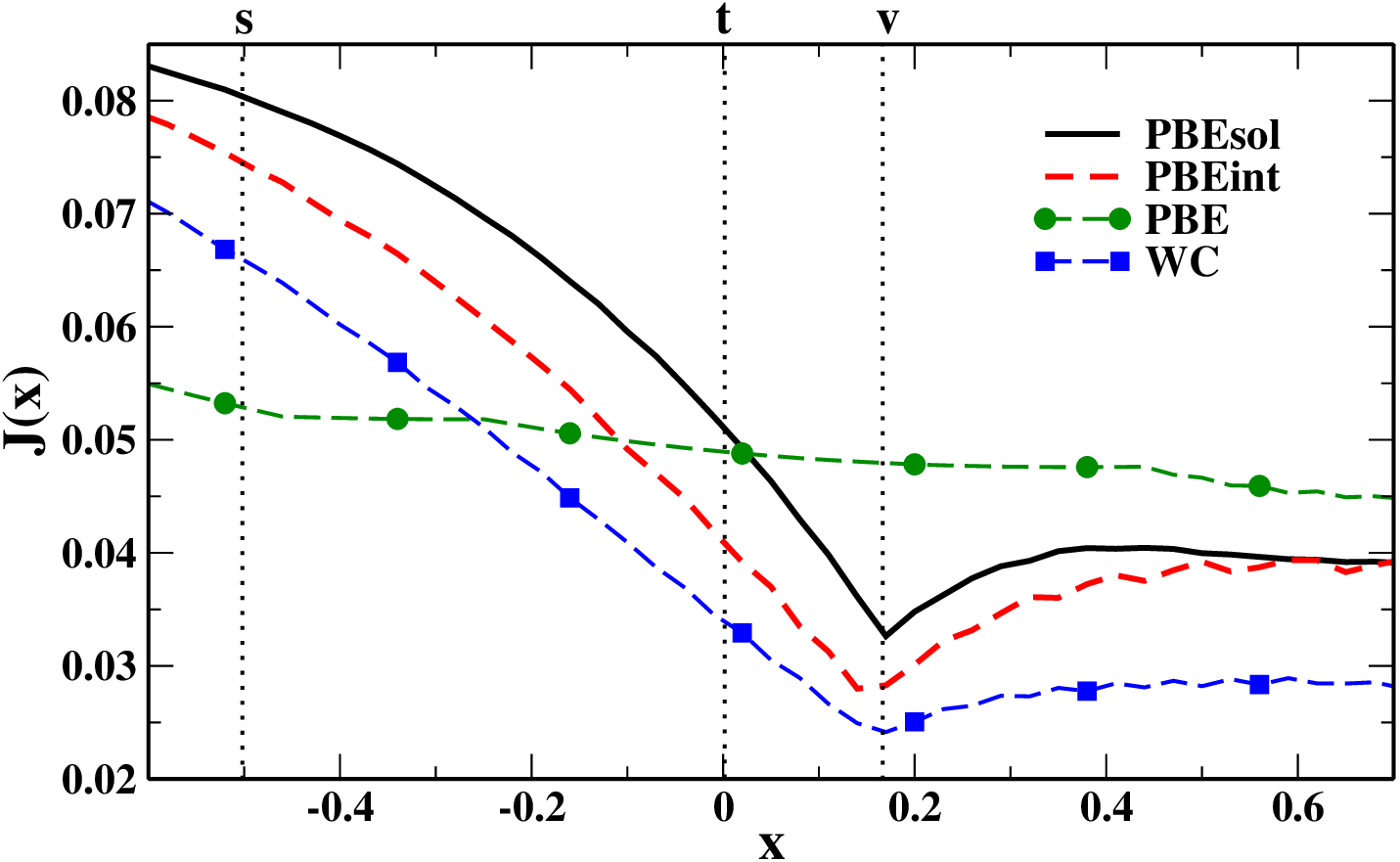}
\caption{$J(x)$ (defined in Eq. (\ref{e16})) for PBE, PBEsol, PBEint, and 
Wu-Cohen GGA \cite{WC}. }
\label{f1}  
\end{figure}
%
The minimization procedure for $x$ is also indicated in 
Fig. \ref{f1}, where we plot the function
\begin{equation}\label{e16}
J(x)=\min_\alpha I[\alpha,x]\ ,
\end{equation}
for several popular functionals. 
The functionals constructed for solids and interfaces (PBEsol, PBEint, and Wu-Cohen \cite{WC})
show a similar trend with $x$:
the value of $J$ is large for $x=-0.5$, that represents the spin 
correction $e^{-\alpha s^3}$; for $x=0$, corresponding to $f=e^{-\alpha t^3}$,
i.e. the previous ansatz used in zPBEsol and zPBEint, it is significantly 
smaller and close to the absolute minimum; at $x=1/6$, all three curves
minimize the function $J(x)$.
For functionals more accurate for atomization energies, as PBE, the 
spin-dependent correction 
is not important (and thus $J(x)$ is almost constant), as shown in Fig. \ref{f1}.

\begin{figure}
\includegraphics[width=\columnwidth]{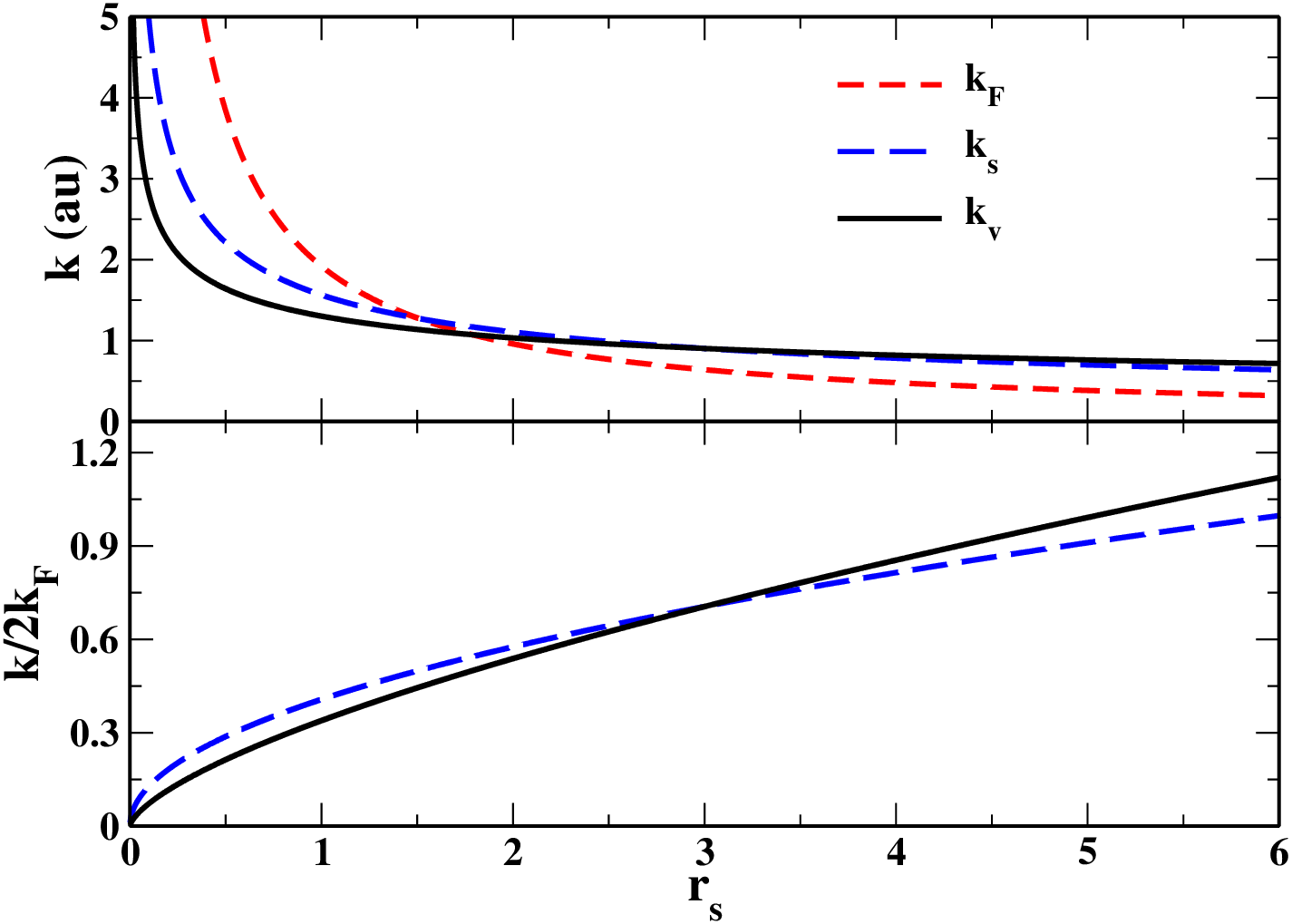}    
\caption{\label{kplot} Dependence of wave vectors $k_F$, $k_s$,
and $k_v$ on the Seitz radius $r_s$ (a.u.). The bottom panel reports the
ratio of wavevectors $k_s$ and
$k_v$ with $2k_F$.}
\end{figure}

We note at this point that the  ansatz of Eq. (\ref{e15}) used to 
define the density parameter $v$ and the value $x=1/6$, obtained from
the minimization procedure, are important outcomes that go
beyond the optimization of the parameters of the model given by 
Eq. (\ref{e14}). 
Thus, the minimization of $J(x)$ allows us to define  
the density parameter
\begin{equation}\label{e17}
v = t\phi \left(\frac{r_s}{3}\right)^{-1/6} = \frac{|\nabla n|}{2k_vn},
\end{equation}
with $k_v=2 (3/(4\pi^4))^{1/18}n^{1/9}$, being indeed a physically 
meaningful quantity to describe density variations in valence and
bonding regions.
In fact, unlike $s$ and $t$ which were derived
in the slowly-varying density limit, $v$ was derived from the statistical
constraint involving the one-electron densities, which are characterized by 
a rapidly-varying regime and connected to cohesive energies of different systems. 
Moreover, Fig. \ref{kplot} shows that, in agreement with the indications
of wave vector analysis on jellium surface energies \cite{LP1,CPP}, $k_v$ and
the ratio $k_v/2k_F$ provide small values of the wave vector
in the high-density regions (small $r_s$) and relatively large values
($\sim 1$) in valence and bonding regions ($r_s\gtrsim 3$), thus granting a 
good sensitivity for density variations in a broad range of $r_s$ values, 
and a superior description (e.g. with respect to $k_s$) of high 
wave-vector contributions in low-density valence and bonding regions.
%
%

Wavevector analyses for the correlation energies have been 
performed only for very simple systems, as jellium surfaces \cite{LP1,CPP,PBEsolhole,JS}
and Hooke's atom \cite{BPL}.
For such systems, the random phase approximation (RPA) fails at large 
wavevectors ($k/2k_F\geq 0.8$), whereas the semilocal functionals performs good
(see Fig. 9 of Ref. \onlinecite{PBEsolhole}), but 
there is still room for improvement \cite{BPL}. We also recall that 
the Langreth and Mehl GGA \cite{LM}, which was one of the first  
semilocal functionals beyond the gradient expansion, is based on the 
wavevector decomposition of the XC energy. 
$k_v$, which was found from one-electron analysis (see Fig. 1), and 
agrees well with the Thomas-Fermi wave vector ($k_s$) (see Fig. 2), 
is more sensitive in rapidly-varying regions.
Thus, the density parameter $v$ and the local wave vector $k_v$,
measuring the range of density variations important for bonding properties,
can be regarded as important ingredients in functional development and further 
work shall be planned to fully assess their importance in this context
(beyond the present spin-dependent correction).

The factor of 3 that enters the $v$ expression was chosen only for
convenience, i.e. for any $x$, the parameter $\alpha$ which minimizes $I[GGA]$
to be between 0 and 2.
In fact, the minimization of $J(x)$ can define the density parameter
$v$ up to a constant, that later may be fixed from a theory of
rapidly-varying density regions.
Nevertheless, the here defined $k_v$ performs similarly with $k_s$ 
in a broad range of density regimes (see Fig. 2). 

The parameter $\omega$ could not be fixed from a minimization
of the statistical-entropy-like function, because for fully
spin-polarized systems the ansatz of Eq. (\ref{e14}) does not depend on
$\omega$. We must therefore introduce additional constraints based on
partially polarized model systems. To this end we consider the
uniformly spin-polarized Gaussian densities
\begin{equation}\label{e18}
\UP=\frac{1+\zeta}{2}\frac{e^{-r^2}}{\pi^{3/2}},\;\;\;
\DN=\frac{1-\zeta}{2}\frac{e^{-r^2}}{\pi^{3/2}},
\end{equation}
that are models for the electronic localization in Wigner crystals 
and in bonding regions of organic molecules \cite{perdew04},
and were already employed to fix constraints in the development of
meta-GGA functionals \cite{TPSS,revTPSS,JS,loc}.
The following conditions are imposed:
\begin{eqnarray}\label{e19}
E_{xc}^{zvGGA}\approx E_{xc}^{GGA}  \, \text{for}\, \zeta\leq 0.3, \quad \nonumber \\
E_{xc}^{zvGGA}\approx E_{xc}^{zGGA} \, \text{for}\, \zeta\geq 0.7,
\end{eqnarray}
where the first condition shall guarantee that the spin-dependent correction
doesn't modifies the original functional 
in the core region of spin-polarized atoms  (where  $0 < \zeta \leq 0.3$) 
while the second one requires the GGAs using the new spin-dependent correction
to be the same as the previous zGGAs (i.e. zPBEsol or zPBEint),
whose behavior at large $\zeta$ had been proved to be remarkably good \cite{zeta}.

We find that for both cases
considered here (GGA=PBEint and PBEsol) the optimum value
of $\omega$ is 9/2. Thus, we can finally define two non-empirical 
functionals making use of the spin-dependent correction of Eq. (\ref{e14})
with the density parameter of Eq. (\ref{e17}), having $\omega=9/2$ and 
$\alpha=1.0$ (zvPBEint) and $\alpha=1.8$ (zvPBEsol), respectively.

\begin{figure}
\includegraphics[width=\columnwidth]{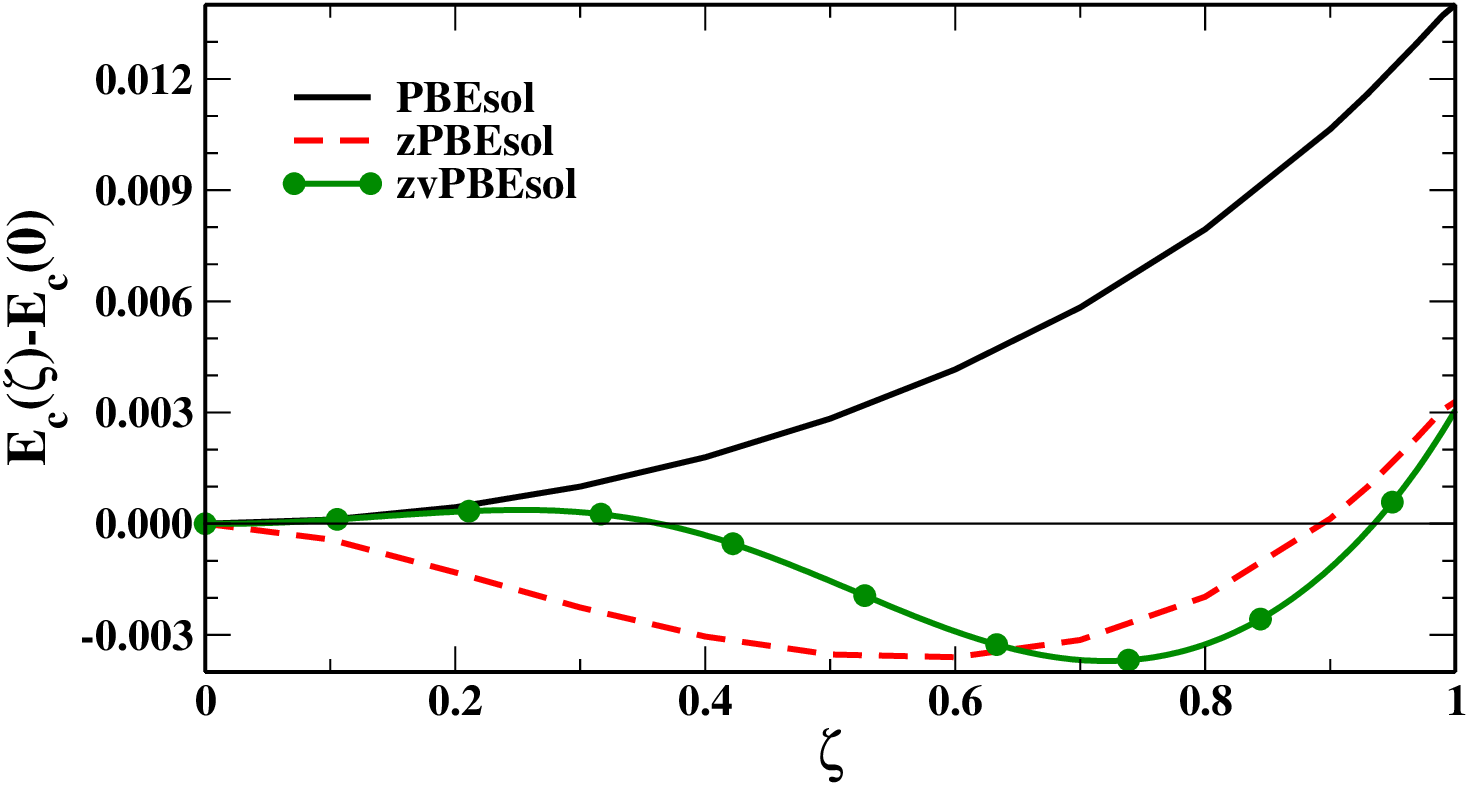}
\caption{\label{f2} $E_c(\zeta)-E_c(\zeta=0)$ (hartree) versus $\zeta$ for 
the one-electron Gaussian densities with uniform spin-polarization $\zeta$ 
(see Eq. (\ref{e18})), for PBEsol, zPBEsol, and zvPBEsol.}
\end{figure}

The plot of $E_c(\zeta)-E_c(\zeta=0)$ versus $\zeta$ for 
the one-electron Gaussian densities with uniform spin-polarization (Eq. 
(\ref{e18})) is shown in Fig. \ref{f2} for PBEsol, zPBEsol, and zvPBEsol.
Similar results (not reported) are found for the PBEint case.
Here we consider the only correlation because the exchange functional is the same
in the considered functional. 
%
%
By construction, zvPBEsol satisfies very well the constraints
of Eq. (\ref{e19}). Note that zPBEsol,
based on the old ansatz shows a pronounced $\zeta$-dependence already for
small values of the spin-polarization.

%
\begin{figure}
\includegraphics[width=\columnwidth]{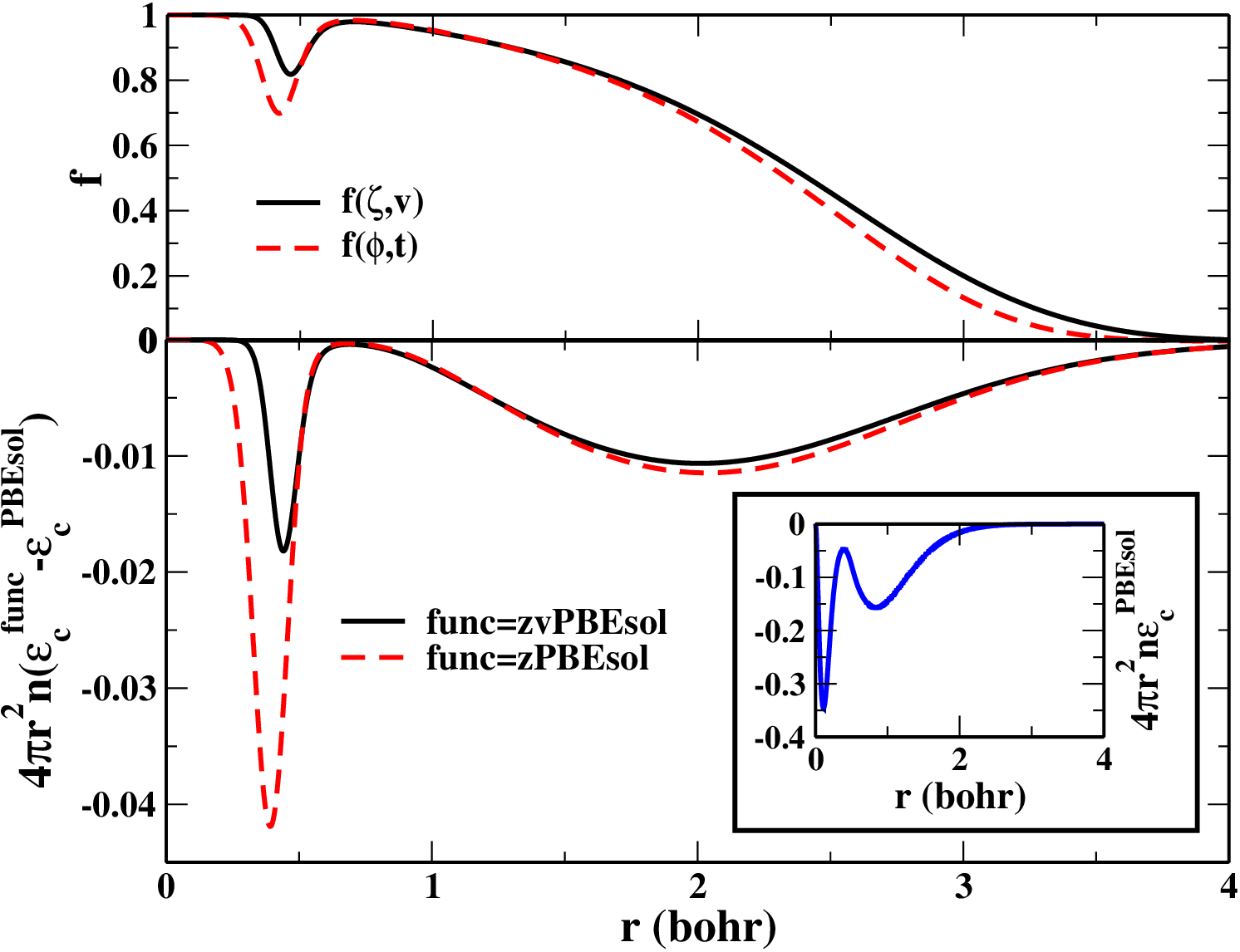}
\caption{\label{fig3} Upper panel: 
Spin-dependent correction 
factors $f(v,\zeta)$ of zPBEsol (Eq. (\ref{e5})) and zvPBEsol (Eq. (\ref{e14})), 
versus radial distance $r$ in N atom;
Lower panel: The deviations from PBEsol of zPBEsol and zvPBEsol,
$4\pi r^2 n (\epsilon_c^{func}-\epsilon_c^{PBEsol})$ (a.u.) versus $r$ in N atom.
In the inset, is shown $4\pi r^2 n \epsilon_c^{PBEsol}$ (a.u.) versus $r$.
}
\end{figure}
%
In Fig. \ref{fig3} 
we also plot the behavior of the PBEsol spin-dependent correction,
 and the deviation from PBEsol of the correlation energy density, 
versus the radial distance in N atom.
As expected, the zvPBEsol improves over zPBEsol in the core region of the
atom, being smooth and very close to the original PBEsol curve, thanks
to its smaller derivative around $\zeta=0$. In the valence and tail regions,
on the other hand, zPBEsol and zvPBEsol perform rather similarly (by construction) and
decrease significantly the correlation with respect to PBEsol.

 Finally, we
  stress that the corrected functionals (zvPBEsol and zvPBEint) fulfill all
  the exact constraints of the original functionals and additionally satisfy the
  new XC constraint for atomization energies.  The Lieb-Oxford bound  \cite{LiebO1,LiebO2,LiebO3}
$E_x\geq E_{xc}\geq 2.27 E_x^{LDA}$, is locally recovered by the exchange part 
of the functionals, and is globally recovered by the full XC functionals 
(in the same manner as PBE and PBEsol), because 
$E_{xc}^{PBEsol}\gtapprox  E_{xc}^{zvPBEsol} \gtapprox E_{xc}^{PBE}$
(see Table S12 in the supplementary material).

\section{Results of zvPBEsol and zvPBEint calculations}
\label{sec4}
In this section we report the results of the application of the
zvPBEint and zvPBEsol functionals to the calculation of the
atomization energies and other properties of several systems. In detail, 
we consider the atomization energies of organic molecules 
(AE6 \cite{AE6} and W4 \cite{GG1} tests),
transition-metal complexes (TM10AE test \cite{FP1,zeta}), 
and small gold clusters 
(AUnAE test \cite{gold}). Note that for all these systems,
especially for those of the TM10AE test, the atomization/interaction
energies are obtained as energy differences between species with different
spin-polarization. Thus, they constitute a natural test for the functionals
developed in the previous section which carry a spin-dependent correction
aiming at improving cohesive and interaction energies.

In addition to the above mentioned properties, we considered also 
different properties of spin-polarized systems in order to verify, in analogy
with Ref. \onlinecite{zeta}, that the spin-dependent correction does not
worsen significantly properties other than cohesive ones, for which it was
designed. Namely, we considered
the binding energies  of small organic molecules to gold microclusters 
(small-interface (SI7) test \cite{PBEint0}), as a model for spin-polarized hybrid interfaces, 
equilibrium bond lengths of  open-shell organic molecules (BL9 test \cite{zeta}) and transition-metal
complexes (TM10BL test \cite{zeta}), 
kinetic properties of small organic reactions (K9 test \cite{lynch03}),  
the spin-states calculations of Table II of Ref. \onlinecite{zeta},  collected 
in the $\Delta E_S$ test (iron \cite{Pierloot} and cobalt \cite{Radon} complexes),
 and the mean absolute errors on the total XC energies of several spin-polarized
atoms and molecules \cite{supp} (${\rm E_{xc}^{atoms}}$ and ${\rm E_{xc}^{mol}}$).

The results are reported in Table \ref{tab1} as well as
in Ref. \onlinecite{supp}.
All calculations were performed with a development version of the 
TURBOMOLE program package \cite{turbomole} using the 
def2-TZVPP basis-set \cite{basis} and full-self-consistent densities.
%
 %
\begin{table*}
 \begin{center}
 \caption{\label{tab1} Mean absolute errors (MAEs) for different tests.
(The results of the $\Delta E_S$ 
are expressed as mean absolute
 relative errors).  For each group of functionals (type-sol and type-int)
 the best result is highlighted in bold. The PBE and APBE results are
 also shown for reference. Full results are reported in Ref. \onlinecite{supp}.}
 \begin{ruledtabular}

\begin{scriptsize}

 \begin{tabular}{lcrrrrrrrrrr}
 Test     & units & PBEsol & zPBEsol &zvPBEsol& $\;\;$  &PBEint & zPBEint &
 zvPBEint& $\;\;$  & PBE & APBE \\
 \hline
 \multicolumn{11}{c}{Atomization and binding energies}\\
 AE6 (organic molecules)  &kcal/mol  & 34.90 & 15.72 & \bf{14.21} &  & 24.78 &
 14.62 & \bf{14.06} & & 14.50 & 7.93 \\
 W4 (organic molecules)   &kcal/mol  & 21.45 & \bf{12.64} & 13.72 &  & 15.57 &
 {\bf 11.59} & 12.32 & & 10.75 & 8.56 \\
 TM10AE (transition metals) &kcal/mol & 18.29 & 10.70 & \bf{9.24} &  & 15.58 &
 10.63 & \bf{8.50} & & 13.47 & 8.69 \\
 AUnAE (gold clusters) & kcal/mol &  4.37 &  \bf{1.97} & 2.43 &  & 2.22 & 1.01 &
 \bf{0.85} & & 0.31 & 1.78 \\
 SI7 (hybrid interfaces) & kcal/mol & 3.92 & \bf{2.88} & 3.28 & & 2.80 & 2.72 &
 \bf{2.50} & & 3.69 & 5.80 \\
 \multicolumn{11}{c}{Other properties}\\
 BL9 (organic molecules) & m\AA & \bf{15} & 16 & \bf{15} &  & 17 & \bf{15} & 16
 & & 15 & 15 \\
 TM10BL (transition metals) & m\AA & 19 & \bf{17} & 18 & & 18 & \bf{17} &
 \bf{17} & & 23 & 26 \\
 K9 (kinetics) &kcal/mol & 10.83 &  \bf{9.90} & 10.42 & & 9.17 & \bf{8.73} & 9.03
 & & 7.51 & 6.58 \\
 $\Delta E_S$ (spin states) &\% & 102 & \bf{74} & 87 & & 90 &
 \bf{72} & 81 & & 79 & 76 \\
${\rm E_{xc}^{atoms}}$ (atoms)
                                           & mH & 427 & {\bf 414} & 417 && 394 & {\bf 385} & 387 && 98 & 24 \\  
${\rm E_{xc}^{mol}}$ (organic molecules)
                                           & mH & 718 & {\bf 710} & 714 && 661 & {\bf 657} & 658 && 146 & 83 \\
 \end{tabular}
\end{scriptsize}
 
\end{ruledtabular}
 \end{center}
 \end{table*}
 %
%

Inspection of Table \ref{tab1} shows that all functionals bearing the spin-dependent
correction (zPBEsol, zvPBEsol, zPBEint, and zvPBEint) systematically and 
significantly improve over the original functionals, for all tests
concerning atomization and interaction energies. On the other hand, z- and zv-type
functionals perform rather similarly, although generally a small improvement is 
obtained with the zv-construction with respect to the corresponding z-type
functionals (except for the W4 test). We highlight also that zvPBEsol and
zvPBEint have a similar performance as PBE for organic molecules atomization
energies, while zvPBEint is among the best GGAs for the TM10AE test,
the gold clusters
(see Fig. 2 of Ref. \onlinecite{gold}), and the hybrid interfaces \cite{PBEint0}.
Finally, we recall that both the z- and the zv-functionals preserve
for the spin-unpolarized cases the performance of the original PBEsol
and PBEint functionals, so that they are very accurate for 
properties like lattice constants of paramagnetic solids \cite{mk}, 
geometries of closed-shell molecules \cite{mk} and description of large metal 
clusters \cite{gold}.

Concerning the non-cohesive properties of open-shell systems (equilibrium
bond lengths, kinetic properties, energy differences between spin states,  and
absolute XC energies of atoms and molecules)
we see that the z- and zv-functionals 
have similar accuracy and are often better than the original functional. 
Concerning the comparison between the z- and the zv-functionals, the former are often better, 
but the differences with respect zv-functional are almost negligible. 
Thus we can conclude that  our spin-dependent correction,
aimed at improving atomization energies, does not introduce significant artifacts
that might alter the description of the spin-polarized systems.

In particular, for the $\Delta E_S$ test (energy differences between spin states)
all GGAs perform badly, with a slightly better performance of z-type
functionals, that in fact is due to their (undesirable) behavior in 
the core of atoms. This test is a hard case for any conventional DFT
approach as it is related with problems with the description of a multireference
character by the KS reference system and static correlation issues \cite{cramer09}.
The spin-dependent correction for atomization energies cannot be expected
therefore to provide any relevant improvement for this property.
We also recall that for relative energies of different spin-states of metal complexes the PBE
exchange is known to provide a very poor performance, while much better
results are achieved by employing the PBE correlation functional in conjunction
with the OPTX \cite{handy01} exchange \cite{swart09_1,swart09_2}.  

Thus, we believe that the non-empirical zvPBEint, which is accurate for energetical and 
structural properties of
paramagnetic bulk solids, hybrid interfaces, molecules, as well for
surface energies of semi-infinite jellium and
of simple metals \cite{JS}, can have a broad applicability.

\section{Spin-dependent correction for meta-GGA dynamical correlation functionals}
\label{sec5}
In this last section we consider the possibility to apply the ansatz
of Eq. (\ref{e14}) also beyond the GGA level.
In this case however it will not be related to the atomization energies
through the information-entropy-like function of Eq. (\ref{e2}), because
the meta-GGAs are one-electron-self-correlation free and no information can be gained
from fully polarized one-electron densities.
Instead, we will show that the ansatz of Eq. (\ref{e14}) can be usefully employed
to improve the compatibility of a meta-GGA dynamical correlation functional
with EXX. For simplicity we consider the Hartree-Fock non-local exchange: in Ref. 
\onlinecite{loc} we showed that 
similar results can be obtained using the local KS EXX.
We tested the TPPS
meta-GGA correlation functional \cite{TPSS}, and the recently developed
TPSSloc meta-GGA correlation functional \cite{loc}, which has the same form as 
TPSS, but uses for its construction instead of the PBE GGA correlation, the
PBEloc GGA functional that has the correlation parameter
\begin{equation}\label{e20}
\beta(r_s,t)=0.0375+0.08\;t^2 (1-e^{-r_s^2})\ ,
\end{equation}
which ensures a stronger localization of the correlation energy density,
thus granting a better compatibility with exact exchange \cite{loc}.

The idea of the present work is to improve the compatibility of TPSS/TPSSloc
with exact exchange by using the spin-dependent correction 
\begin{equation}\label{e21}
E_c^{zvMGGA}=\int d\R \; n\; e^{-\alpha v^3 |\zeta|^\omega}\;\epsilon^{MGGA}_c\ ,
\end{equation}
to provide a further
localization of the correlation energy density for spin-polarized densities.
 Here the same 
spin-dependent correction factor of Eq. (\ref{e14}) can be used, as also in this
case we need to modify the original correlation only 
in rapidly-varying spin-polarized  density regime ($\zeta \ne 0$ and $v > 1$), thus
preserving all the other important exact  conditions of the original functional.
The rationale beyond this choice is the fact that from a physical point of view
the localization of spin-polarized densities corresponds to a reduction of the
effective range of same-spin correlation contributions $\epsilon_c[\UP,0]$ and
$\epsilon_c[0,\DN]$, while for opposite-spin contributions 
$\epsilon_c[\UP,\DN]-\epsilon_c[\UP,0]-\epsilon_c[0,\DN]$
an increase of the effective range is obtained. Thus, we can
use the ansatz of Eq. (\ref{e21}) to
$(i)$ fine tune the relative ranges of the same- and opposite-spin 
correlation contributions, that in TPSS (and PBE) were determined from 
LDA and/or jellium constraints \cite{WP1,TPSS}, so probably overestimate 
the same-spin range \cite{becke87} with respect to the opposite-spin one;
$(ii)$ cut the longer-range part of same-spin contributions that, when exact exchange
is used, are no more needed to compensate for the too short range of semilocal
exchange.
  
To build the spin-dependent correction factor and find appropriate values
for the parameters $\alpha$ and $\omega$, we use the uniformly 
spin-polarized Gaussian densities of Eq. (\ref{e18}) and impose the following 
constraints:

(i) At small relative spin-polarizations we must recover the original functional,
hence we must require
\begin{equation}\label{e22}
\frac{dE_c^{zvMGGA}}{d\zeta} \Big|_{\zeta\leq 0.3}\approx 0\ .
\end{equation}
This condition, in analogy with the GGA case assures that in core regions,
where closed-shell configurations dominate and $\zeta$ is small,
the functional mimics the original behavior of the uncorrected functional.
Similar constrains has been used for the (rev)TPSS functionals \cite{TPSS,revTPSS,JS}.
 
(ii) For arbitrary values of the relative spin-polarization the XC functional
must approach as close as possible the $\zeta$-dependence of the
exact XC functional, $E_{xc}(\zeta)$. The latter is known to be constant for one-electron 
densities and for
different values of $\zeta$ \cite{cohen08,Cohen2}, i.e. 
\begin{equation}\label{e23_2}
E_{xc}(\zeta)=E_{xc}(1)=E_x(1),
\end{equation}
where $E_x(1)$ is the exchange energy for the one-electron Gaussian 
density (the correlation energy vanishes for
one-electron systems).

Separating the spin-independent part of the exchange and correlation
energies ($E_x$ and $E_c$ respectively) from their spin-scaling factors we can rewrite Eq. 
(\ref{e23_2}) as:
\begin{equation}
E_x(0)f(\zeta) + E_c(0)g(\zeta) = E_x(0) f(1)\ .
\end{equation}
As the exact spin-scaling factor for the exchange (for one electron systems) is:
\begin{equation}\label{e23}
f(\zeta) = \frac{(1+\zeta)^2+(1-\zeta)^2}{2}\ ,
\end{equation}
the spin-scaling factor for the correlation (for one electron system) must be:
\begin{equation}\label{e24}
g(\zeta)= \left[2-\frac{(1+\zeta)^2+(1-\zeta)^2}{2}\right] = 1-\zeta^2\ .
\end{equation}
Therefore, we impose the constraint
\begin{equation}\label{e24_2}
E_c^{zvMGGA}(\zeta) \rightarrow E_c^{zvMGGA}(0)g(\zeta),\;\;\;\rm{at}\;\;\; \zeta\geq 0.7 .
\end{equation}
In this way in fact the $\zeta$-dependence of the overall EXX+C functional 
(approximately) preserves the exact spin-behavior, at $\zeta\geq 0.7$, while 
performs as the original functional at $\zeta\leq 0.3$.
We note also that Eq. (\ref{e24_2}) implies 
that 
$E_c^{zvMGGA}(1)=E_c^{MGGA}(1)=0$.

\begin{figure}
\includegraphics[width=\columnwidth]{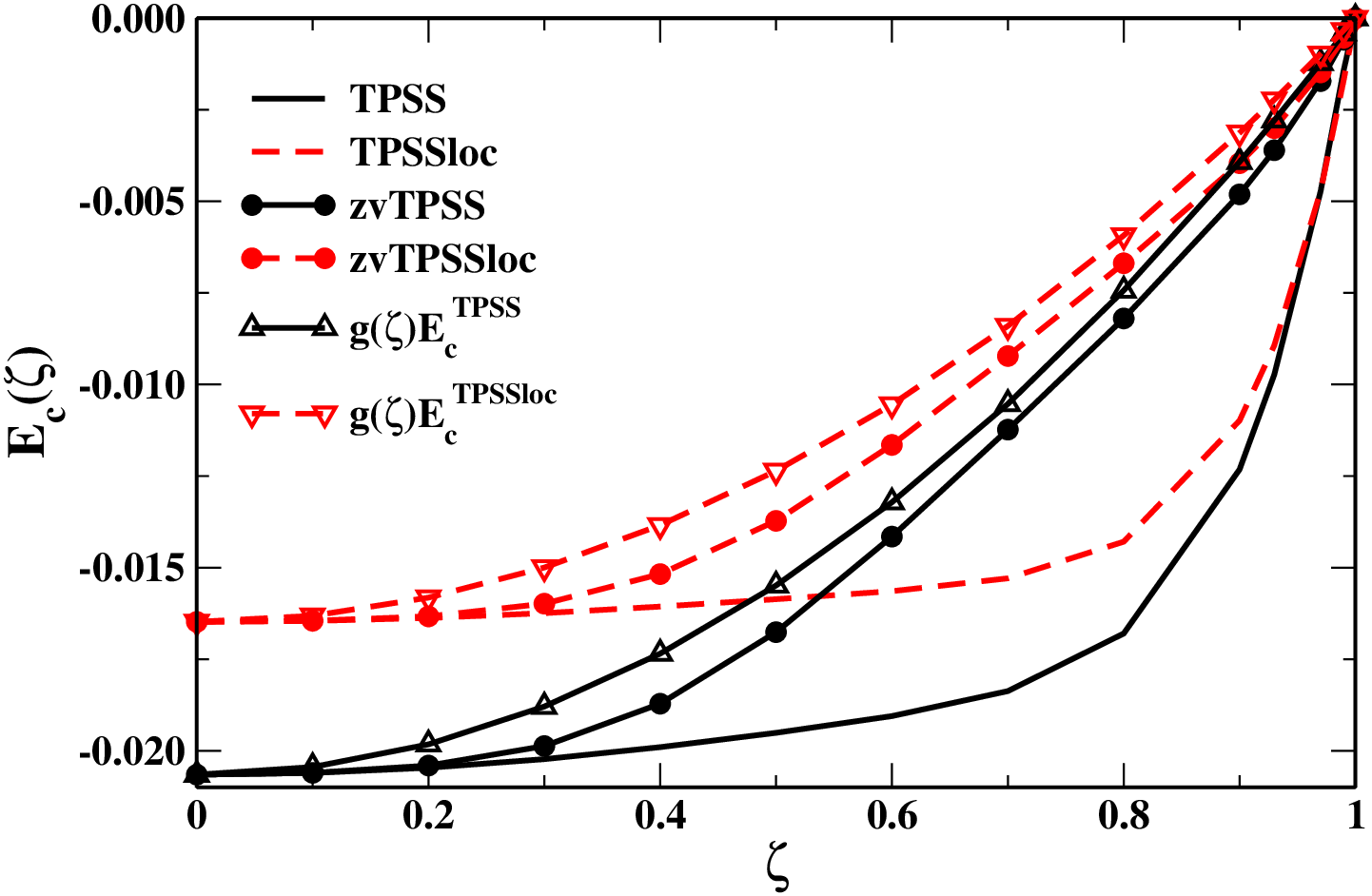}
\caption{ $E_c(\zeta)$ (hartree) versus $\zeta$ of the one-electron Gaussian 
density with uniform spin-polarization $\zeta$ (see Eq. (\ref{e18})), 
for TPSS, TPSSloc, zvTPSS, zvTPSSloc, and ideal $E_c(z)$ of Eq. (\ref{e24}).}
\label{f5}
\end{figure}

By imposing conditions (i) and (ii) we find $\omega=9/2$ and
$\alpha=8$ for meta-GGA=TPSSloc, $\alpha=6$ for meta-GGA=TPSS.
The resulting correlation functionals, to be used with full EXX, can be named 
zvTPSSloc and zvTPSS, respectively.
In Fig. \ref{f5} we show how well the zvTPSSloc and zvTPSS dynamical correlation 
functionals satisfy constraints (i) and (ii) in contrast to the 
original TPSS and TPSSloc functionals, which satisfy constrain (i), but  whose 
$\zeta$-dependence differs 
significantly from the desired one for $\zeta\geq 0.4$. 

%
\begin{figure}
\includegraphics[width=\columnwidth]{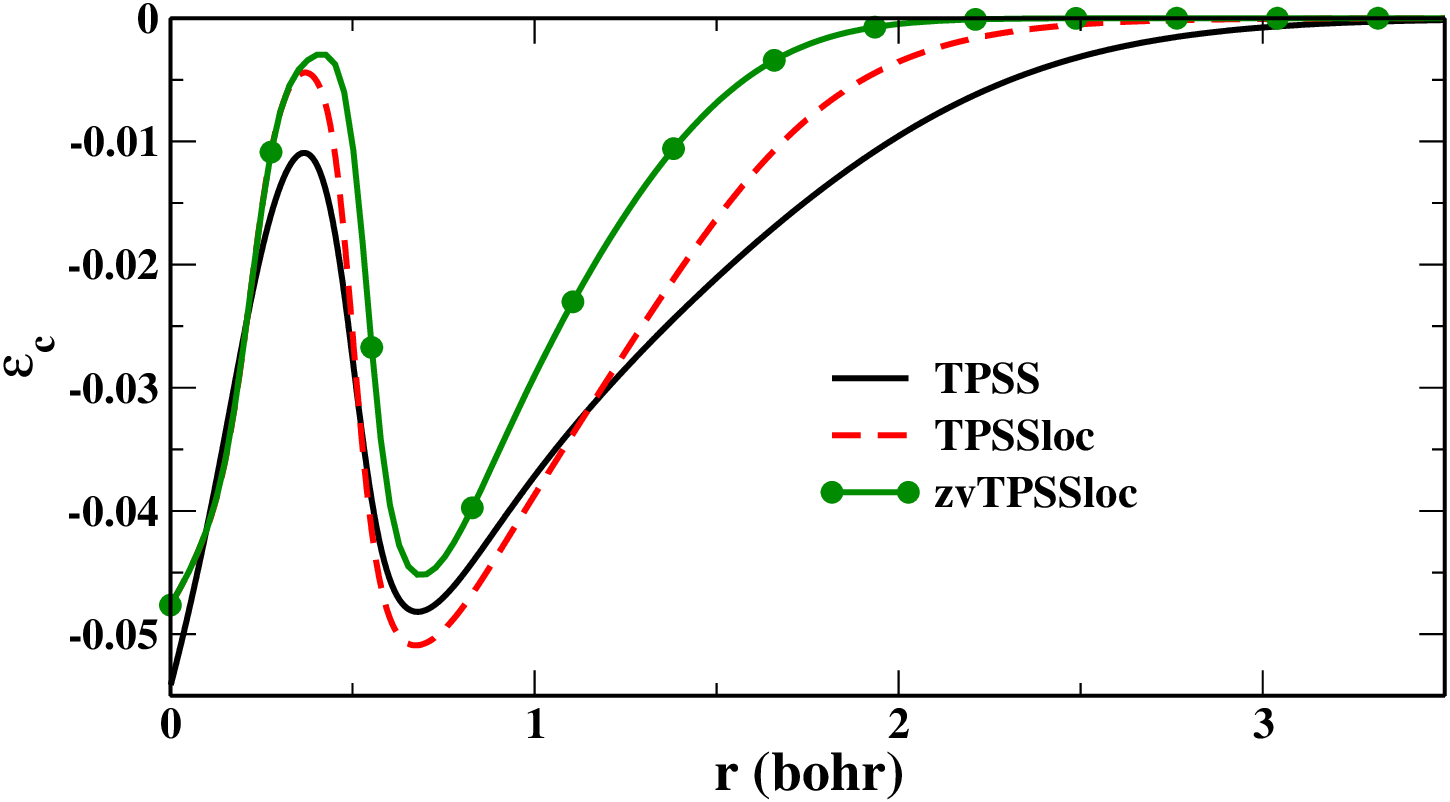}
\caption{ $\epsilon_c$ (a.u.) versus the radial distance $r$ for the N atom 
as obtained from the TPSS, TPSSloc, and zvTPSSloc functionals.}
\label{f6}
\end{figure}

%
In Fig. \ref{f6} we display instead a plot of the TPPS, TPSSloc, and
zvTPSSloc correlation energy densities of the N atom.
%
%
The figure shows that, as required, the zv-correction does not have any 
effect inside the atomic core and only produces a further localization 
for the valence (spin-polarized) density, showing the correspondence of 
the imposed constraints for one-electron uniformly-polarized Gaussian densities
with the physical requirements for the functional.

At this point, we need to explain the differences between the spin-dependent
corrections at
XC GGA level, and at EXX + meta-GGA level. The exchange functionals constructed
for solids and hybrid interfaces (xPBEsol and xPBEint), have  localized holes \cite{PBEsolhole}
and thus they are not accurate for atoms, overestimating the atomic exchange energies.
So, the spin-dependent correction, based on the H, G, and C ensemble, improves the XC
energies of one-electron systems, by making a necessary delocalization of the
spin-dependent correlation hole \cite{zeta}. On the other hand, at the EXX+meta-GGA level,
one electron densities are exactly described, and because the EXX hole is delocalized,
the dynamical correlation part should be more localized.

To test the zvTPSS and zvTPSSloc functionals we applied them to compute 
atomization energies (AE6 test \cite{AE6}), barrier heights
(BH6 test \cite{lynch03}) and reaction kinetic properties 
(K9 test\cite{lynch03}) of small organic systems. More extensive tests
on larger systems are instead not considered, because
in our treatment we do not account for non-dynamical correlation, so that 
only energy properties of simple small systems at equilibrium geometry 
can be computed at present. We recall in addition that for spin-unpolarized
systems the  zvTPSS and zvTPSSloc perform exactly as the original
TPSS and TPSSloc functionals, so we did not consider tests involving only
spin-unpolarized species (e.g. non-bonded interactions).

The results are reported in Table \ref{tab3}.
%
\begin{table}[b]
\begin{center}
\caption{\label{tab3} Mean absolute errors (kcal/mol) on atomization energies 
and kinetics of small organic systems, as resulting from calculations 
using Hartree-Fock exchange and different meta-GGA correlation functionals. 
The best value for each line is shown in boldface. 
Full results are reported in Ref. \onlinecite{supp}.}
\begin{ruledtabular}
\begin{tabular}{lrrrr}
Test set & TPSS & TPSSloc & zvTPSS & zvTPSSloc\\
\hline
AE6 & 29.1    & 25.5 & 20.6 & \bf{17.3}   \\
BH6 & 4.7 & 3.9 & 4.6 & \bf{3.4} \\
K9  & 5.6    & 4.3 & 4.3 & \bf{3.6} \\
\end{tabular}
\end{ruledtabular}
\end{center}
\end{table}
%
From an inspection of the table appears clearly that the zv-corrected correlation
functionals have a significantly higher compatibility with exact exchange and
provide much improved results with respect to the original TPSS and TPSSloc functionals.
In particular, the zvTPSSloc functional, joining the spatial localization of the
correlation energy density and the spin-dependent correction, provides
in all cases the
smallest errors, with MAEs of atomization energies comparable with those of PBE
and very small errors for barrier heights and the K9 test.

\section{Conclusions}
\label{sec6}
In conclusion, we have discussed in detail the GGA spin-dependent
correction of Ref. \onlinecite{zeta}, for atomization energies. 
From an analysis the one-electron statistical ensemble (H, G, and C), we have 
found a density parameter $v\propto |\nabla n|/n^{10/9}$ for the valence and tail density regions, 
whose wave vector $k_v\propto n^{1/9}$, seems the next term is a wavevector series, where 
the first two terms are $k_F\propto n^{1/3}$ (compatible with exchange), and $k_s\propto n^{1/6}$
(compatible with Yukawa interaction \cite{PBE} in the slowly-varying
high-density limit). 

By using the density parameter $v$, we propose a simpler GGA spin-dependent
correction, constructed to be relevant only in the valence and tail regions, where 
the spin polarization is important. The here proposed GGAs (zvPBEsol and zvPBEint)
systematically improve the atomization and binding energies of molecular 
systems with respect PBEsol and PBEint, 
preserving the accuracy of the original functionals for other properties, and thus achieving 
a broader applicability. 

The results of Table I, show that the zvPBEint
functional performs remarkably well for energetical and structural properties of 
transition metals, being more accurate than the APBE GGA reference, and thus 
they can be considered in applications of transition metal chemistry, where 
popular functionals (including hybrids) give a modest behavior. 

The z- and zv- functionals, perform similarly for atomization 
and binding energies, but better results are found from zv- type, especially 
for the PBEint case. On the other hand, for other properties, the zv- functionals 
are closer to the original ones (by construction).
The construction of non-empirical GGAs with
broad applicability (i.e., accurate for molecules, bulk solids and surfaces),
is of great theoretical and practical interest. Recent work proved that
non-empirical GGAs can not be both accurate for atoms and solids \cite{PCSB}, despite
highly empirical GGA functionals demonstrated a good accuracy for a broad
range of problems \cite{Truhlar10}. Thus, the development of non-empirical 
GGA functionals with broad
applicability, is a theoretical challenge in DFT. In this work we have shown
that the spin-dependent correction (applied to satisfy the statistical
condition for atomization energies, and other physical conditions derived from model systems), 
may be one path through solving this challenge.

We have also applied the spin-dependent correction to meta-GGAs dynamical correlation functionals,
showing a significantly better compatibility with the EXX. In fact, the 
zv-corrected meta-GGAs improve more than 30\% over the original ones, for 
atomization energies. Thus, we believe that the zvTPSSloc meta-GGA correlation 
functional, that is accurate for jellium surfaces, and Hooke's atom at any frequency \cite{loc},
can be a good starting point in the development of more accurate (and non-empirical)
hyper-GGAs \cite{hyper1,hyper2}.

Finally, in this paper, we have shown that the spin-dependent correction of Ref. \onlinecite{zeta},
which was simplified and improved in this work, is a powerful tool at GGA and meta-GGA levels. 
Thus, we believe that it can be used also to functionals that describe the dimensional crossover 
(from 3D to 2D) of the XC energy \cite{q2D}, in order to improve the atomization and 
binding energies of molecular systems under a 2D confinement.

\textbf{Acknowledgments.} This work was partially funded by the ERC Starting
Grant FP7 Project DEDOM (No. 207441). We thank TURBOMOLE GmbH for the TURBOMOLE
program package.

\end{document}